\documentclass[twocolumn,aps,prl,showpacs,groupedaddresst,superscriptaddress]{revtex4-1}
\usepackage[latin9]{inputenc}
\usepackage{color}
\usepackage{amsmath}
\usepackage{amssymb}
\usepackage{graphicx}
\usepackage[unicode=true]
 {hyperref}

\makeatletter

\usepackage{dcolumn}
\usepackage{bm}
\usepackage{changes}
\usepackage{amsfonts}

\makeatother

\begin{document}
\title{Testing Critical Slowing Down as a Bifurcation Indicator \\
 in a Low-dissipation Dynamical System}
\author{M. Marconi}
\affiliation{Universit\'e C\^ote d\'~Azur, Institut de Physique de
Nice, CNRS - UMR 7010, Sophia Antipolis, France}
\author{C. M\'etayer}
\affiliation{Universit\'e de la Nouvelle Cal\'edonie, ISEA, BP R4 -
98851 Noum\'ea Cedex, Nouvelle Cal\'edonie}
\author{A. Acquaviva}
\affiliation{Universit\'e C\^ote d\'~Azur, Institut de Physique de
Nice, CNRS - UMR 7010, Sophia Antipolis, France}
\author{J.M. Boyer}
\affiliation{Universit\'e de la Nouvelle Cal\'edonie, ISEA, BP R4 -
98851 Noum\'ea Cedex, Nouvelle Cal\'edonie}
\author{A. Gomel}
\affiliation{Universidad de Buenos Aires, Departamento de Fisica,
Intendente Guiraldes 2160, CABA, Buenos Aires, Argentina}
\author{T. Quiniou}
\affiliation{Universit\'e de la Nouvelle Cal\'edonie, ISEA, BP R4 -
98851 Noum\'ea Cedex, Nouvelle Cal\'edonie}
\author{C. Masoller}
\affiliation{Departamento de Fisica, Universitat Politecnica de
Catalunya, St Nebridi 22, Barcelona 08222, Spain.}
\email[Corresponding author: ]{cristina.masoller@upc.edu}
\author{M. Giudici}
\affiliation{Universit\'e C\^ote d\'~Azur, Institut de Physique de
Nice, CNRS - UMR 7010, Sophia Antipolis, France}
\author{J.R. Tredicce}
\affiliation{Universit\'e de la Nouvelle Cal\'edonie, ISEA, BP R4 -
98851 Noum\'ea Cedex, Nouvelle Cal\'edonie} \affiliation{Universidad
de Buenos Aires, Departamento de Fisica, Intendente Guiraldes 2160,
CABA, Buenos Aires, Argentina}

\date{\today}
\begin{abstract}
We study a two-dimensional low-dissipation dynamical system with a
control parameter that is swept linearly in time across a transcritical
bifurcation. We investigate the relaxation time of a perturbation
applied to a variable of the system and we show that critical slowing
down may occur at a parameter value well above the bifurcation point.
We test experimentally the occurrence of critical slowing down by
applying a perturbation to the accessible control parameter and we
find that this perturbation leaves the system behavior unaltered,
thus providing no useful information on the occurrence of critical
slowing down. The theoretical analysis reveals
the reasons why these tests fail in predicting an incoming bifurcation. 
\end{abstract}

\maketitle
There has always been a special interest in trying to predict transitions,
crisis, and catastrophic events. 
Today, a huge amount of research is devoted to determine good indicators
applicable on time series obtained from real systems that may anticipate
a change in its behavior or, in the language of dynamical systems,
a bifurcation \cite{Scheffer2009,Scheffer2012,Malik2014}. This is
particularly relevant in disciplines like medicine, biology, atmospheric
science, ecology, sociology, economy, where these predictions may
avoid a disaster or, at least, they may be useful to prepare the system
to a behavioral change. For example, it has been conjectured that
the advent of an epilepsy attack is the result of a phase transition
\cite{Litt2001,McSharry2003}, that climate on earth is actually very
close to a tipping point \cite{Lenton2008}, that extremely intense
pulses in lasers may result from a bifurcation of a chaotic attractor
\cite{Granese2016,Metayer2014} and that evolutive specialization
in ecology \cite{Thompson1998,Dakos2012} is also the consequence
of a bifurcation. We may say that any behavioral change in
a real system is connected to the existence of a bifurcation in the
corresponding dynamical system and that the prediction of these changes
depends on the possibility of establishing reliable indicators alerting
of the incoming bifurcation. 

A well-established indicator which follows from the definition of
bifurcation is known as ``critical slowing down" (CSD). When the
system approaches a bifurcation, its relaxation time after a perturbation
grows asymptotically and this divergence is referred to as CSD \cite{Mori}. {\textcolor{black}{CSD is often associated to an increase of the variance and of the autocorrelation of a variable of the system~\cite{variance}. Nevertheless, it has been observed that these indicators are not always reliable for alerting on an incoming bifurcation
\cite{Burthe2016,Guttal2016}. }}

On the other hand, real systems evolve towards a bifurcation because
one or more parameters are changing in time. For example, the level of 
CO$_{2}$ in the atmosphere is an evolving parameter {\textcolor{black}{that may lead the earth's climate system to a bifurcation~\cite{co2}.}}

In this paper we address the fundamental question whether CSD is always
a good indicator of an incoming bifurcation in a system where a parameter
is linearly changing in time. By definition, CSD can be identified
by perturbing the dynamical system. Unfortunately, in real systems,
this perturbation cannot be implemented in the variables but rather
in the parameters that are accessible in the experiments. Hence, a
second question that we address here is whether a perturbation of
an evolving control parameter can be a reliable probe for testing
the occurrence of a bifurcation. We answer to these questions by presenting
a real system with a time swept parameter where CSD appears only after
the bifurcation has already occurred, hence when it might be too late
to reverse the change in behavior. Furthermore, we show that a perturbation in the accessible control parameter is unable to provide any indication, nor on the occurrence
of CSD nor on the bifurcation crossing.

We begin by considering a simple two-dimensional dynamical system describing a class-B laser \cite{Metayer2014},
\begin{eqnarray}
dS/dt=-S(1-N),\;\nonumber \\
dN/dt=-\gamma(N-A+SN).\;\ \label{eq:one}
\end{eqnarray}
Here $S$ is proportional to the light intensity and $N$ to the
atomic population inversion; $A$ is proportional to the pump and
$\gamma$ is 
the ratio between the decay rates of the population inversion ($\gamma_{p}$) and the intensity ($\gamma_{i}$) . The time $t$ is normalized to $\gamma_{i}$. This dynamical system exhibits a transcritical bifurcation at $A=1$. For
$A<1$ the solution $(S,N)=(0,A)$ is stable. For $A>1$ the solution
$(S,N)=(A-1,1)$ is stable. We vary the pump parameter $A$ with a
triangular ramp of speed $b$, {\textcolor{black}{always smaller than the decay rate}}
of the variables of the system: 
\begin{eqnarray}
A(t)=A_{0}+bt\:\text{ for }\:t\leq t_{0}\;\nonumber \\
A(t)=A_{0}+bt_{0}-b(t-t_{0})\:\text{ for }\:t_{0}\leq t\leq2t_{0}\;\label{eq:two}
\end{eqnarray}
{here $A_{0}$ is the initial value of the pump and $t_{0}$ is the
duration of the ramp-up and of the ramp-down.} 
\begin{figure}
\includegraphics[width=0.8\columnwidth]{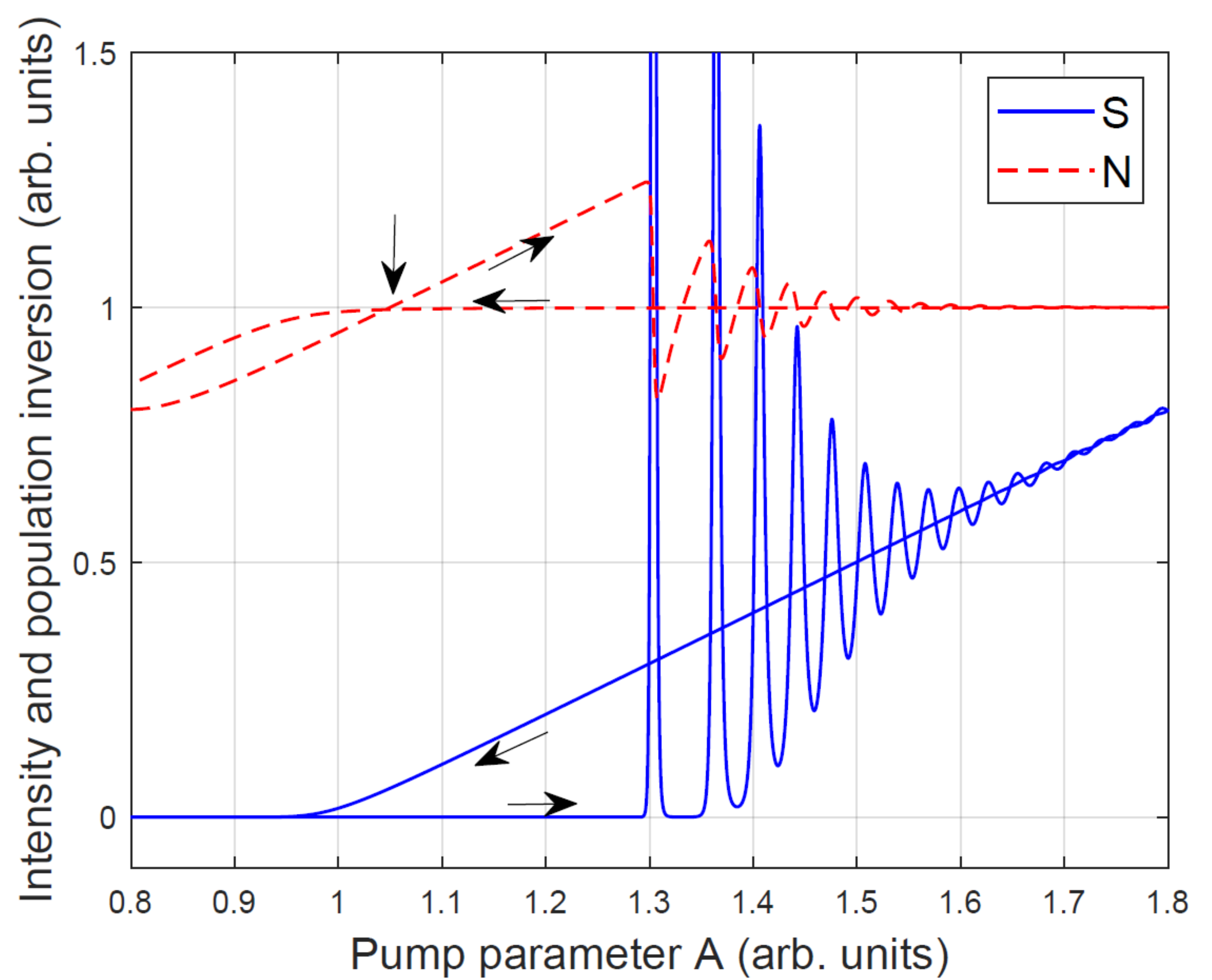} \caption{\label{fig:one} Intensity, $S$, and population inversion, $N$,
as a function of the pump, $A$. The pump is swept at a velocity $b=0.0005$;
$\gamma=0.01$. The initial conditions are $S_{0}=0.001,N_{0}=0.8,A_{0}=0.8$.
The arrows indicate how $S$ and $N$ evolve as the pump is increased
and then decreased. The vertical arrow shows that $N=1$ is reached
when $A=1.05$.}
\end{figure}

The evolution of the laser intensity $S$ as a function of the pump
parameter $A$ is plotted in Fig.~\ref{fig:one}. $S$ grows significantly
at a value well beyond the bifurcation point $A=1$. This delayed
reaction of a laser when the pump is swept across the threshold was
studied theoretically \cite{Mandel1984} and experimentally \cite{Scharpf1987,Tredicce2004}. {\textcolor{black}{
Critical slowing down was put in evidence in \cite{Tredicce2004} by measuring the asymptotical growth of this delay as a function of the speed of the pump change. For a vanishing speed this delay diverges, thus revealing the presence of CSD at the laser threshold. }}
As shown in Fig. Fig.~\ref{fig:one}, the intensity remains close to zero on a large interval during which
the pump continues to grow beyond the threshold value. Hence the system
accumulates energy, which is suddenly released leading to a spike-like
variation of the intensity. If the system is under damped ($\gamma<1$),
as in the situation considered in Fig.~\ref{fig:one}, relaxation
oscillations occur until an asymptotic solution is reached, where
the intensity follows the pump. This behavior is typical
of class-B lasers {\textcolor{black}{\cite{Arecchi1984,bimberg},}} such as semiconductor or solid
state lasers.

\begin{figure}
\includegraphics[width=1\columnwidth]{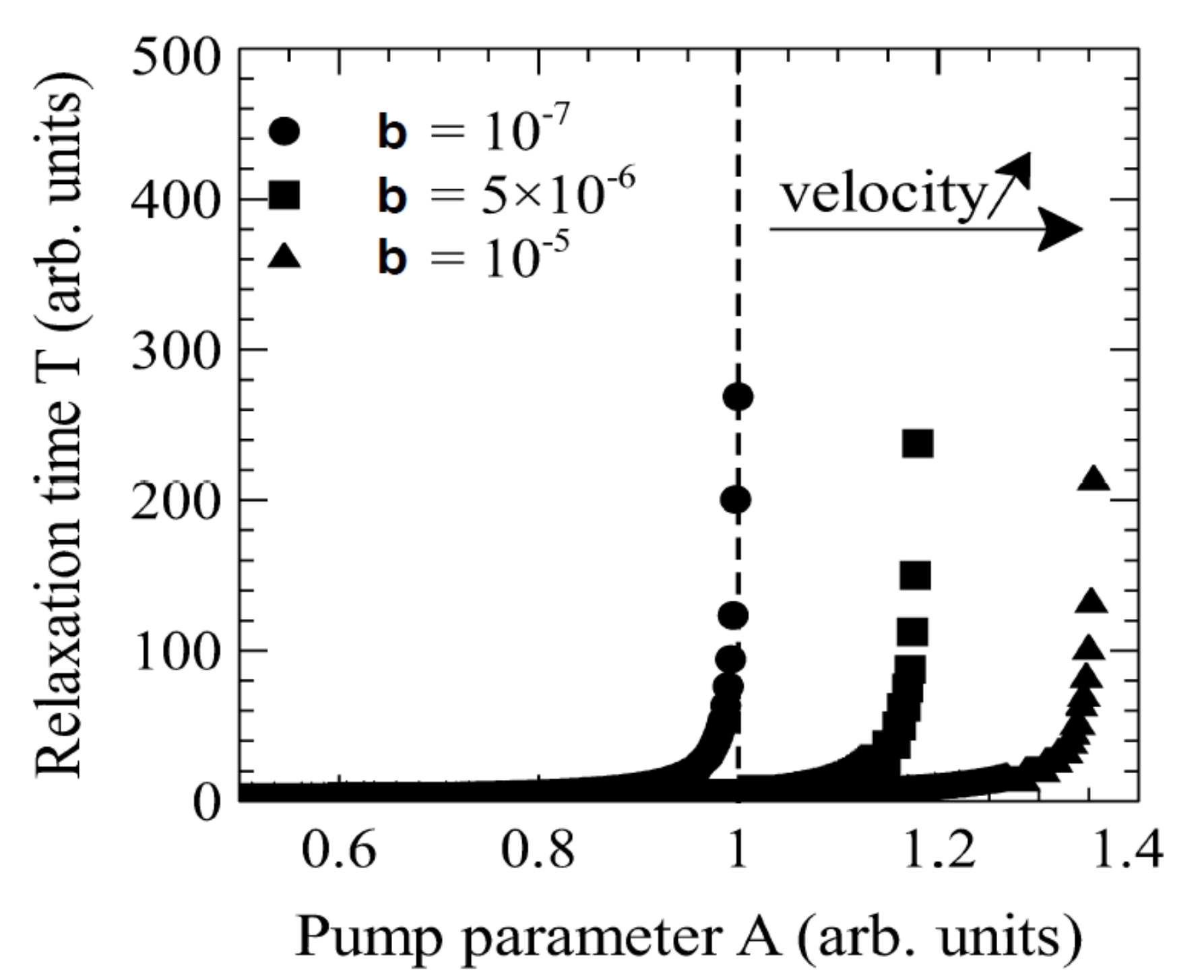} \caption{\label{fig:two} Relaxation time $T$ of the laser intensity $S$
after this variable was perturbed by a short pulse as a function of
the pump value $A$. Three 
values of the sweep velocity $b$ are considered, while $\gamma=2.7\times10^{-5}$
is kept constant. The occurrence of CSD is marked by the asymptotic
growth of $T$. For increasing $b$, CSD takes place at values
of $A$ larger than the bifurcation point ($A=1$, indicated with
a dashed line).}
\end{figure}

Here, we test numerically the occurrence of CSD in this system through a
small perturbation $\Delta S$ in the laser intensity at different
values of the pump parameter $A$. We measure the time taken by the
perturbation to decrease to 1/e of its initial value. Our results
are plotted in Fig.~\ref{fig:two}. 
{\textcolor{black}{We notice that, for slow ramps, the relaxation time diverges
at the bifurcation point, i.e. when we approach $A=1$, as demonstrated in \cite{Tredicce2004}.}} However,
for larger values of $b$, CSD does not take place at $A=1$ but at
a higher value of the pump parameter, which increases with $b$.

In order to explain these observations, we note that, when $S\approx0$
(i.e., before the laser turns on), the evolution of a perturbation
$\epsilon$ of the intensity is governed by:
\begin{equation}
\frac{d\epsilon}{dt}=\epsilon(N-1)\label{eq:three}
\end{equation}

The relaxation time diverges when $d\epsilon/dt=0$, so when $N$ becomes
equal to 1. Importantly, the amplitude of the perturbation has no
effect whatsoever during this stage ($S\approx0$). Because the pump
parameter grows linearly in time (as indicated in Eq.~\eqref{eq:two}
for $t\le t_{0}$), the equation governing the evolution of $N$ is:
\begin{equation}
\frac{dN}{dt}=-\gamma(N-A_{0}-bt),\label{eq:four}
\end{equation}
{whose solution is $N(t)=A_{0}+bt-{b\left(1-e^{-t\gamma}\right)}/{\gamma}$}~{\cite{aclaracion}}.
Then, by imposing $N=1$, we can calculate analytically the critical
pump value ($A_{c}$) at which the relaxation time diverges and CSD
takes place:
\begin{equation}
A_{c}=1+\frac{b}{\gamma}+\frac{b}{\gamma}W\left[-e^{-\frac{\gamma}{b}(1-A_{0})-1}\right],\label{eq:five}
\end{equation}
with $W$ being the Lambert $w$ function. Hence, for the parameters used
in Figs.~\ref{fig:one} and~\ref{fig:two}, $A_{c}$ depends mainly on the ratio $b/\gamma$. As this ratio increases, $A_{c}$ grows above the value at which the
bifurcation takes place ($A=1$). The effect of $A_{0}$ on $A_{c}$
is negligible provided that $A_{0}<1-b/\gamma$. In agreement with
this analytical estimation, in the simulations, using the parameters
of Fig.~\ref{fig:two}, one finds $A_{c}=1.004$, $1.18$ and $1.37$
for $b=1\times10^{-7}$, $5\times10^{-6}$ and $1\times10^{-5}$ respectively,
while in Fig.~\ref{fig:one}, $N=1$ is reached when $A=1.05$, {indicated
by the vertical arrow}.

Therefore, we have identified a system with a time-swept parameter
in which CSD takes place well beyond the bifurcation point, contradicting
the common belief that CSD is an indicator of an upcoming bifurcation.
Two ingredients are needed for the dynamical system to behave in this
counterintuitive manner: a fast sweeping rate of the parameter and low dissipation.
The system considered is a 
class-B laser where $\gamma$ (the ratio between the decay rates of the population inversion, $\gamma_{p}$, and the intensity, $\gamma_{i}$) is significantly
smaller than one. 

To meet this requirement we perform experiments with a diode-pumped solid state lasers (SSL) Nd:YVO4 emitting at 1.060 $\mu$m. In this laser $\gamma_{p}$ is of the
order of $2\times10^{4}$, while $\gamma_{i}$ is $5\times10^{9}$,
leading to $\gamma\approx4\times10^{-6}$ (see Supplementary Material).

SSL threshold is observed for a bias current $J$ of the diode pump
$J=J_{th}=147\,$mA.  The diode pump laser can be modulated by a triangular
ramp applied to its bias current, hence sweeping linearly the pump
intensity from a zero level ($J=88\,$mA, which corresponds to the diode pump\textcolor{black}{{} threshold) up to 1.4 times the threshold
value of the SSL ($J=1.4\times J_{th}=208\,$mA). The laser package
is thermally stabilized in a temperature range where the SSL emits
on the same single longitudinal mode in the whole swept pump range.
The input and output signals (the bias current of the diode pump and
the intensity of the SSL respectively) are monitored on a digital
oscilloscope. 
The ramp duration can be varied from 0.05~s to 0.25~ms. The speed of the fastest ramp, as defined in Eq. \ref{fig:two}, is $b=1.12\times10^{-6}$, hence $b/\gamma=0.28$
(see Supplementary Material). According to Eq.~(\ref{eq:five}),
this upper value of $b/\gamma$, together with the possibility of
controlling experimentally $b$, makes this laser an ideal system
to test the prediction of this equation. Unfortunately, as in the
majority of real systems, it is not possible to perturb directly the
laser variables ($S,N$) to probe the occurrence of CSD. However,
real systems can be perturbed through their control parameters and
we may check their influence on the variables}.

\begin{figure}
\includegraphics[width=1\columnwidth]{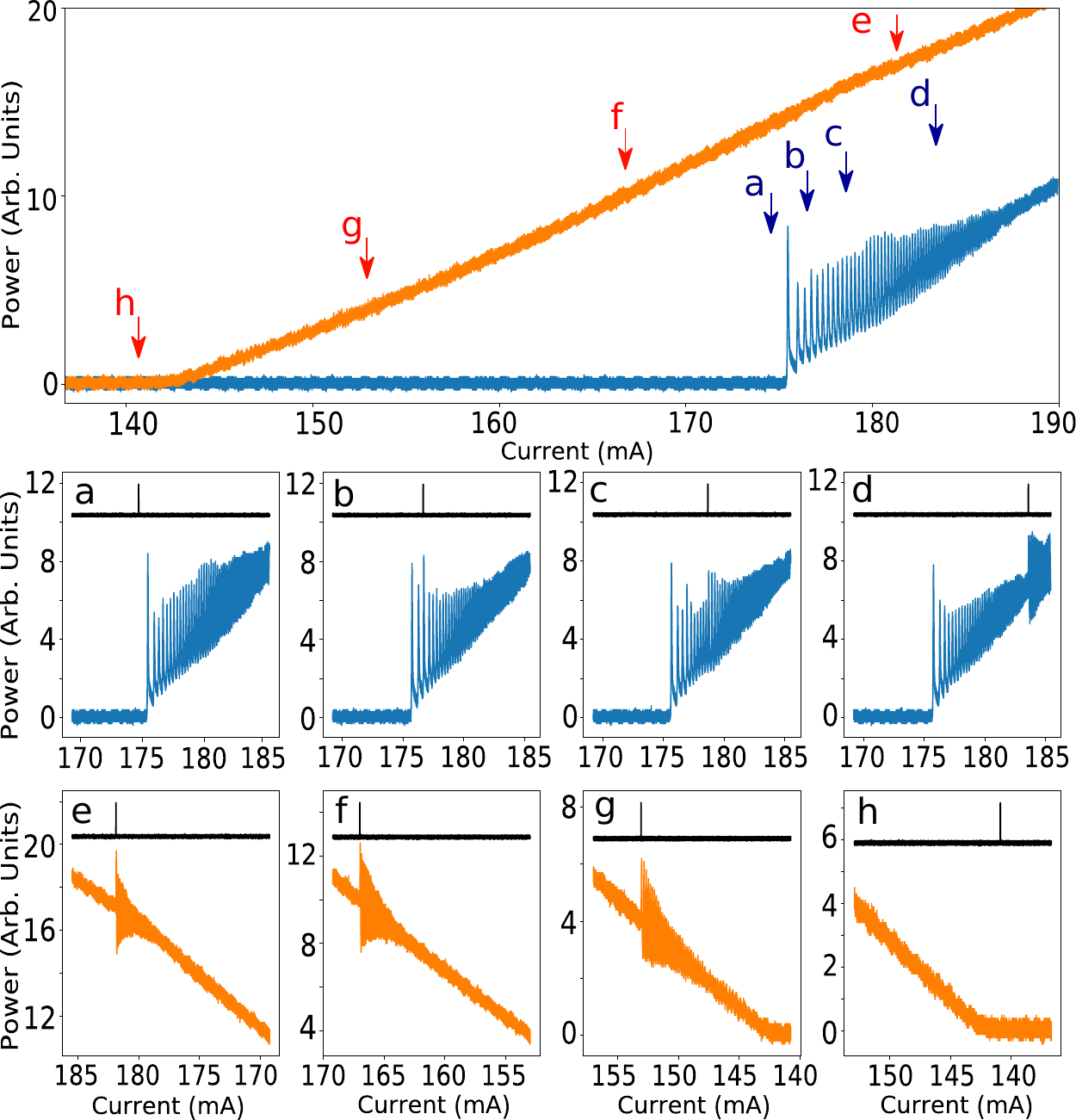} \caption{\label{fig:three} Upper Panel: Laser output intensity as a function
of the bias current of the diode pump that is swept in time by using
a triangular ramp (see text for details). The intensity trace obtained
during the positive (negative) slope of the ramp is displayed in blue
(orange). Panels a-h): A short pulse (1 $\mu s$ width and 40 mA height)
is superimposed over the pump ramp at the positions indicated by the
arrows on the upper panel. The effect of each perturbation on the
laser intensity is shown in the corresponding panels.}
\end{figure}

The SSL laser intensity output versus the time-varying
pump level is shown in the upper panel of Fig.~\ref{fig:three} for
a modulation rate of 100 Hz. We notice, in agreement with Fig.~\ref{fig:one},
that the laser intensity grows significantly only when the pump current
is well above the threshold (the intensity spike occurs at $J=J_{on}=$175.5
mA, which corresponds to $1.2J_{th}$). The first lasing peak is followed
by damped relaxation oscillations whose frequency increases as the
pump increases. The delayed response of the SSL in terms of the pump
level has been investigated for different speed of the bias current
ramp $b$ and we have observed that it follows the $b^{-1}$ law predicted
in~\cite{Mandel1984}. This delay is almost absent when ramping down
the pump current and the laser switches off at $J\approx J_{th}$.
The difference between the pump value at which the laser starts to
emit ($J=J_{on})$ and the pump value at which it switches off ($J\approx J_{th}$)
leads to the well-known dynamical hysteresis of Fig.~\ref{fig:three},
which has also been observed in~\cite{Tredicce2004}.

For the low modulation rate used in Fig.~\ref{fig:three}
we estimate $b=5.6\times10^{-8}$ and $b/\gamma=0.014$ (see Supplementary
Material). Hence, according to Eq.~(\ref{eq:five}), CSD is expected
to occur very close to the SSL threshold $(J=J_{th})$. We test the
occurrence of CSD by adding to the bias current of the diode pump
a perturbation pulse which is synchronous with the current ramp. By
varying the phase of the two signals we can place the perturbation
at arbitrary positions of the ramp and analyze the response in the
intensity variable. We apply a pulse of 40 mA with a duration of 0.5~$\mu s$  at {\textcolor{black}{full-width half-maximum (FWHM)}}. We superimpose it to the pump ramp and in Fig.
\ref{fig:three}, we show the most relevant positions, marked by the
arrows in the upper panel. 

The evolution of the perturbation depends
clearly on its position on the ramp, as shown by the panels a) to
h). When the pump current is ramped up and the pulse is applied before
the laser emits the first spike and switches on (panel a), the intensity
is not affected and it remains at the level of the experimental noise.
This is observed for any position of the perturbation in the interval
$0<J<J_{on}=1.2J_{th}$. If the perturbation is applied after the
laser has switched on, the pulse may enhance the next relaxation oscillation
peak (panel b). Instead, when it is applied between two relaxation
oscillation peaks, it will decrease the amplitude of the next relaxation
oscillation. In any case, the perturbation pulse induces a new transient
in the relaxation process started after the laser switch-on. This
can be seen in panels c) and d) where the perturbation is applied
when the laser oscillations are significantly damped. One can notice
that the relaxation is faster when the perturbation is applied closer
to the top of the ramp, i.e. at the maximum value of the pump current.
When the pump current is ramped down, the evolution of the perturbation
is not interacting with another relaxation process and therefore,
it is more clearly visualized: during the ramp down the perturbation
induces damped relaxation oscillations. As the perturbation is applied
closer to the bifurcation point where the laser switches off, the
damping time increases while the relaxation oscillations decreases
(panels e,f,g). Finally, after the laser switches off, the perturbation
does not induce any reaction on the intensity variable (h). 

These experimental evidences indicate that no signature
of CSD, nor of the bifurcation crossing at $J=J_{th}$, can be found
by perturbing the pump parameter when the system evolves from the
off-state to the on-state. This surprising behavior is observed
for any speed of the ramp, even for the highest ones, where CSD is expected
to occur well-beyond the threshold value of the SSL. When $b$ is
increased the laser switches on at an increasing pump level (for example,
with a ramp duration of 0.5 ms, $J_{on}=1.3\times J_{th}$ ) according
to the law predicted in~\cite{Mandel1984}, and no effect on the
intensity output is noticed when the pump perturbation is applied
in the interval $0<J<J_{on}$. Instead, the perturbation pulse does
have an effect on the intensity output when the laser is in the on
state. In this case, intensity exhibits a spike followed by damped
relaxation oscillations whose frequency and damping rate decrease as the perturbation is
applied closer to the bifurcation point ($J=J_{th})$. 

\begin{figure}
\textcolor{black}{\includegraphics[width=1\columnwidth]{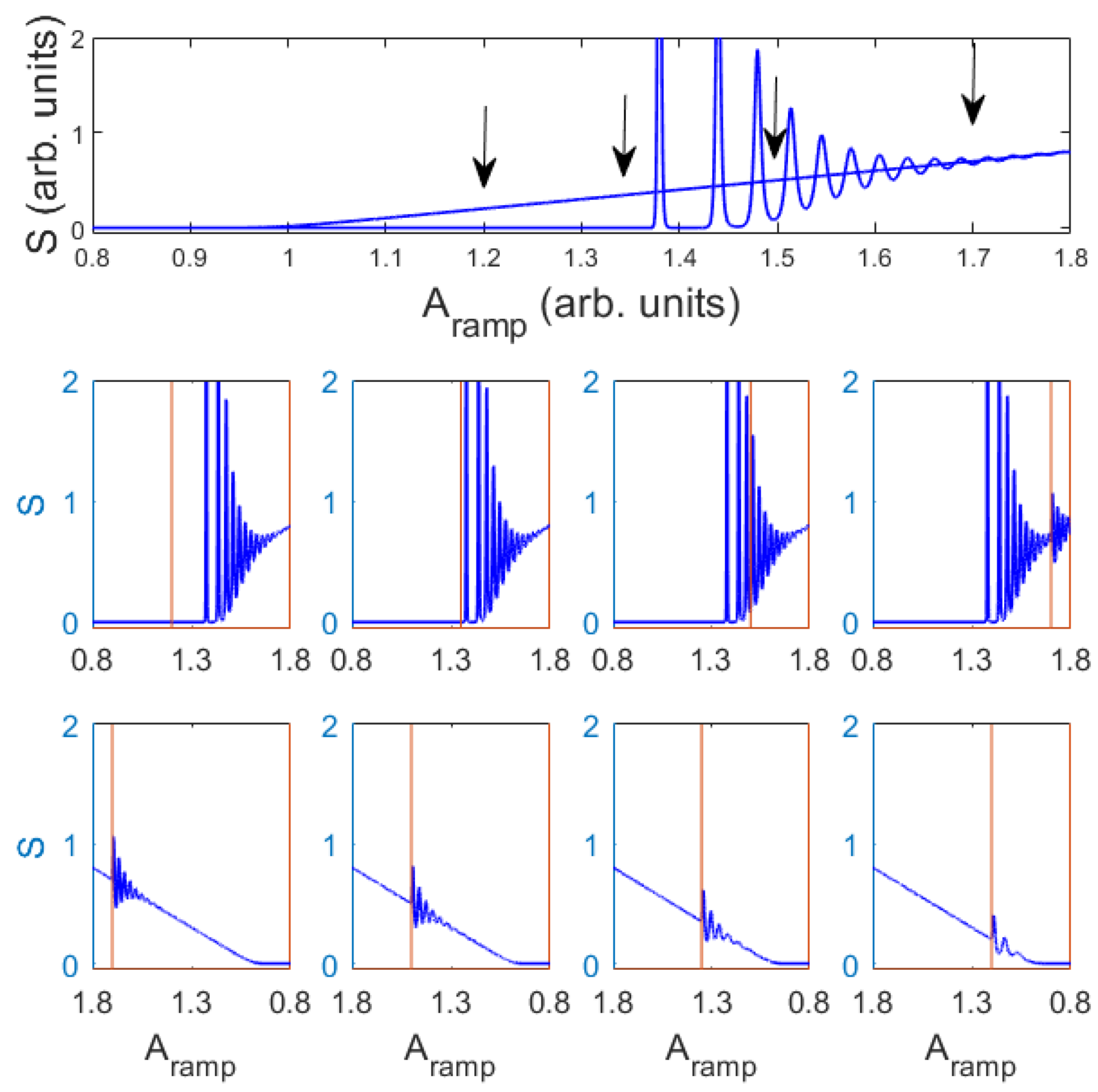}
\caption{\label{fig:four} {Top panel: Intensity dynamics as a function of
the pump. The parameters are as in Fig. 1. The arrows and vertical lines indicate the
position at which we apply a short pulse to the pump (see text for
details). Central (bottom) panels: intensity dynamics when the pulse is applied
during the upward (downward) ramp.}}
}
\end{figure}

In order to understand why perturbing the control
parameter is not a reliable method to probe the occurrence of CSD
in our laser, we have used Eqs. \ref{fig:one} and \ref{fig:two} to analyzed numerically the effect of a short pulse in the pump parameter,
i.e., $A(t)=A_{ramp}(t)+A_{p}(t)$, where $A_{ramp}(t)$ is the triangular
signal described by Eq.~(\ref{eq:two}) and$A_{p}(t)$
is a short rectangular pulse. The results obtained, displayed in
Fig.~\ref{fig:four}, are in very good agreement with the experimental
findings. We remark that for the parameters used in Fig.~\ref{fig:four}
$b/\gamma=0.05$, and therefore, according to Eq.~(\ref{eq:five})
CSD occurs at $A\sim$1.05. Nevertheless, as in the experiments, no
response to the pulse is observed in the laser intensity as long as
the laser is off. Simulations including noise show similar results (see Supplementary
Material).

\textcolor{black}{}

The response of the system to a short perturbation in the pump can
be understood by analyzing the structure of the equations. When the
laser is off, the intensity $S$ vanishes and $N$ must, in principle,
follow the pump parameter $A$. However, being $N$ a slow variable
($\gamma<<1$), it is unable to follow a sudden variation of $A$,
as, for example, when $A$ is perturbed by a short pulse. Hence, the
pump pulse does not affect the value of the variable $N$ which will
just continue to follow the pump ramp and $S$ will remain close to
zero, even if the perturbation pulse is applied when $A>A_{c}$, i.e.
when the pump parameter is beyond the critical point where $N=1$
and CSD occurs. In fact, no response in the $S$ variable to the pump
pulse can be observed before the laser switches on. After the first
laser spike, $S>0$ and this variable will respond directly to a perturbation
pulse in the pump, thus "bypassing" the lowpass filtering of the
variable $N$. In this condition, the relaxation process following
the pump perturbation is observable in the varia\textcolor{black}{ble
$S$, and an effect of the pump pulse on the intensity output can
be measured; however, this occurs }only after the laser has turned
on. 

In conclusion, we have shown that, in a low-dissipation system with
a control parameter that is swept linearly in time, CSD is not always a reliable
indicator of an incoming bifurcation. {\textcolor{black}{By considering a two-dimensional real system featuring a transcritical bifurcation, we have demonstrated that CSD
may occur beyond the bifurcation point, which makes it useless for
alerting of an incoming behavioral change of the system.}}
Moreover, we have shown that a perturbation of an evolving 
parameter might not able to identify CSD: this occurs when the parameter affects directly a slow variable. In this case,
a fast perturbation pulse may leave this variable unchanged and will
have no effect on the system output. {\textcolor{black}{While these results can be generalized to any dynamical system having a dimension $ \ge 2$ bifurcating transcritically, we are currently investigating their extension to other type of bifurcations.}} 

We believe that our results have important impact in environmental
studies, in particular, in ecosystems' dynamics{\textcolor{black}{~\cite{eco1,eco2,cris}}}, because the evolution of populations are often described by coupled nonlinear rate equations
as those considered here, and control parameters such as the amount
of water or food available, can vary in time.
\begin{acknowledgments}
JRT thanks the program ECOS-Sud A14E03 Ev\'enements extr\^emes en dynamique non lin\'eaire  for granting support to this research. MG and MM acknowledge
ANR Blason (ANR-18-CE24-0002). CM acknowledges partial support from
Spanish MINECO (PGC2018-099443-B-I00) and from the program ICREA ACADEMIA
of Generalitat de Catalunya. 
\end{acknowledgments}


\begin{thebibliography}{99}
\bibitem{aclaracion} {If the pump ramp starts well below the threshold, the initial condition of the population inversion is $N(0)=A_0$.}
\end{thebibliography}

\end{document}